# Evaluation of MRI to ultrasound registration methods for brain shift correction: The CuRIOUS2018 Challenge

Yiming Xiao, Member, IEEE, Hassan Rivaz, Member, IEEE, Matthieu Chabanas, Maryse Fortin,
Ines Machado, Yangming Ou, Mattias P. Heinrich, Julia A. Schnabel, Xia Zhong, Andreas Maier,
Wolfgang Wein, Roozbeh Shams, Samuel Kadoury, David Drobny, Marc Modat, and
Ingerid Reinertsen

*Abstract*—**In brain tumor surgery, the quality and safety of the procedure can be impacted by intra-operative tissue deformation, called brain shift. Brain shift can move the surgical targets and other vital structures such as blood vessels, thus invalidating the pre-surgical plan. Intra-operative ultrasound (iUS) is a convenient and cost-effective imaging tool to track brain shift and tumor resection. Accurate image registration techniques that update pre-surgical MRI based on iUS are crucial but challenging. The MICCAI Challenge 2018 for Correction of Brain shift with Intra-Operative UltraSound (CuRIOUS2018) provided a public platform to benchmark MRI-iUS registration algorithms on newly released clinical datasets. In this work, we present the data, setup, evaluation, and results of CuRIOUS 2018, which received 6 fully automated algorithms from leading academic and industrial research groups. All algorithms were first trained with the public RESECT database, and then ranked based on test dataset of 10 additional cases with identical data curation and annotation protocols as the RESECT database. The article compares the results of all participating teams and discusses the insights gained from the challenge, as well as future work.**

## I. Introduction

GLIOMAS are the most common brain tumors in adults, and are categorized into grade I-IV by the World Health Organization (WHO). Low-grade gliomas (LGG, grade I and II) are less aggressive and have slower progression than high-grade gliomas (HGG, grade III and IV), but will eventually undergo malignant transformation into high-grade tumors. Evidences [1, 2] have shown that early tumor resection can effectively improve the patient's survival rate. Image-guidance can be a useful tool to assist the surgeon in obtaining a maximal safe resection of the tumor. Image-guidance based on pre-operative MR images is in routine clinical use worldwide. These systems, however, do not account for the tissue shift and deformations that occur as the resection progresses. Due to brain shift, the surgical target and other vital structures (e.g., blood vessels and ventricles) will be displaced relative to the pre-surgical plan and resulting in inaccurate image-guidance. Multiple factors can contribute to brain shift, including but not limited to drug administration, intracranial pressure change, tissue resection. Often such tissue shift is not directly visible by the surgeon. Both intra-operative ultrasound (iUS) and intra-operative magnetic resonance imaging (iMRI) have been employed to track tissue deformation and surgical progress. Intra-operative US has gained popularity thanks to its low cost, high portability and flexibility. However, limited field of view and challenging image interpretation remain obstacles for widespread use. Together with iUS, automatic image registration algorithms can be used to update the surgical plan based on pre-operative MRI by re-aligning the pre-operative images with intra-operative images and offer more intuitive assessments of the extent of resection.

This manuscript was submitted on xx, revised on xx, and accepted on xx.

Y. Xiao is with the Robarts Research Institute, Western University, London, ON, Canada (e-mail: yxiao286@uwo.ca).

H. Rivaz is with the PERFORM Centre and Department of Electrical and Computer Engineering, Concordia University, Montreal, Canada (e-mail: hrivaz@ece.concordia.ca).

M. Chabanas is with University of Grenoble Aples, Grenoble Institute of Technology, Grenoble, France (e-mail: Matthieu.Chabanas@univ-grenoble-alpes.fr).

M. Fortin is with the Department of Health, Kinesiology & Applied Physiology, Concordia University, Montreal, Canada (e-mail: maryse.fortin@concordia.ca).

I. Machado is with the Department of Radiology, Brigham and Women's Hospital, Harvard Medical School, Boston, MA, USA.

Y. Ou is with the Department of Pediatrics and Radiology, Boston Children's Hospital, Harvard Medical School, Boston, MA, USA.

M.P. Heinrich is with the Institute of Medical Informatics, University of Luebeck, Germany.

J.A. Schnabel is with the School of Biomedical Engineering and Imaging Sciences, King's College London, UK.

X. Zhang and A. Maier are with the Pattern Recognition Lab, Department of Computer Science, Friedrich-Alexander-Universität Erlangen-Nürnberg, Martensstr. 3, 91058 Erlangen, Germany.

R. Shams and S. Kadoury are with the Institute of Biomedical Engineering, Polytechnique Montreal & CHUM Research Centre.

D. Drobny is with the Wellcome/EPSRC Centre for Interventional and Surgical Sciences, University of College London, UK.

D. Drobny and Marc Modat are with the School of Biomedical Engineering & Imaging Sciences, King's College London, King's Health Partners, St Thomas' Hospital, London, SE1 7EH, UK.

W. Wein is with ImFusion GmbH, Munich Germany.

I. Reinertsen is with SINTEF, NO-7465 Trondheim, Norway, (e-mail: Ingerid.Reinertsen@sintef.no).

Previously, a number of algorithms and strategies [3-8] have been developed to address iUS-MRI registration for brain shift correction. They range from new strategies to map image features to similar domains [3, 4] to novel cost function [6, 7], and from different deformation models [5, 6] to improved optimization procedures [8]. However, partially due to the lack of relevant clinical datasets, it has been difficult to directly compare different algorithms, thus potentially slowing the speed of technical translation to benefit surgeons and patients. The MICCAI Challenge 2018 for Correction of Brain shift with Intra-Operative UltraSound (CuRIOUS2018) was launched as the first public platform to benchmark the latest image registration algorithms for the task, and to bring the researchers together to discuss the technical and clinical challenges in iUS-guided brain tumor resection. For the first edition of the challenge, we focused on MRI-iUS registration to correct pre-resection deformation after craniotomy, as it typically sets the tone of brain shift for the rest of the surgery.

The challenge was divided into two phases. In the first phase, the publicly available REtroSpective Evaluation of Cerebral Tumors (RESECT) database [9] was used to train registration algorithms. The participating teams then submitted a description of their methods and results on the training data. Two weeks before the challenge ended, a private testing database curated under the same condition as the training set was supplied to the participants who submitted their results based on the training data. For both training and testing data, the distance between homologous anatomical landmarks across iUS and MRI were used to assess and rank the registration quality. The CuRIOUS2018 challenge received 8 initial submissions [10-17]. Seven teams validated their methods on the testing data, and six participated in the final ranking. The submissions cover a wide variety of approaches, including the latest registration metrics [7, 13], optimization approaches [12], and deep learning techniques [17].

This paper describes the organization, submitted algorithms, and results for the challenge, and further discusses the current challenges and potential future directions of tissue shift correction in US-guided brain tumor surgery.

## II. MATERIALS

Two datasets were included for the training and testing phases of the CuRIOUS2018 challenge. The RESECT database [9] was provided to the participants as the training dataset for development and fine-tuning of the algorithms. The database contains pre-operative MR and pre-resection iUS images from 22 patients who have received LGG resection surgeries at St. Olavs University Hospital, Trondheim, Norway. The testing dataset was comprised of imaging data from 10 additional patients with LGG obtained in the same setting as the RESECT database. The collection and distribution of both datasets were approved by the Regional Committee for Medical and Health Research Ethics of Central Norway, and all patients signed written informed consent.

For both training and testing databases, Gd-enhanced T1w MRI and T2w fluid-attenuated inversion recovery (FLAIR) MRI scans were acquired for each patient before surgery. Five fiducial markers were glued to the patient's head prior to scanning. The T1w and T2w MRIs were rigidly co-registered, and aligned to the patient's head position on the operating table via a fiducial-based image-to-patient registration. The position-tracked 3D iUS scans were acquired with the Sonowand Invite neuronavigation system (Sonowand AS, Trondheim, Norway), with either the 12FLA-L linear transducer or the 12FLA flat linear array transducer for smaller superficial tumors. 3D volumes were reconstructed from the raw iUS data using the built-in proprietary reconstruction method in the Sonowand Invite system, with a reconstruction resolution in the range of 0.14x0.14x0.14 $mm^3$ to 0.24x0.24x0.24 $mm^3$ depending on the probe types and imaging depth. Both ultrasound transducers were factory calibrated and equipped with removable sterilizable reference frames for optical tracking. A Polaris camera (NDI, Waterloo, Canada) built in the Sonowand system was used to obtain the position and pose of the ultrasound probe. Therefore, the iUS volumes reveal tissue position and deformation in the patient's head on the operating table.

TABLE I
DETAILS OF INTRA-MODALITY LANDMARKS FOR EACH PATIENT IN THE TESTING DATASET

| Patient ID | # of landmarks MRI vs. before US | Mean initial distance (range) in mm MRI vs. before US |
|---|---|---|
| 1 | 17 | 15.66 (14.19~16.74) |
| 3 | 17 | 6.36 (3.57~10.23) |
| 4 | 17 | 2.98 (1.17~5.28) |
| 5 | 17 | 13.19 (9.86~17.25) |
| 6 | 18 | 5.52 (4.07~7.24) |
| 7 | 18 | 5.27 (4.28~6.14) |
| 8 | 18 | 3.73 (2.66~5.04) |
| 9 | 17 | 1.80 (0.41~4.15) |
| 10 | 17 | 4.66 (3.76 ~5.74) |
| 12 | 17 | 4.89 (3.58~6.21) |
| mean±sd | 17.3±0.5 | 6.41±4.46 |

The number of landmarks and mean initial Euclidean distances between landmark pairs are shown, and the range (min ~ max) of the distances is shown in parenthesis after the mean value.

Homologous anatomical landmarks manually labeled by two raters (authors YX and MF as *Rater 1* and *2*, respectively) were provided to assess registration quality, using the software 'register' included in the MINC toolkit (http://bic-mni.github.io). Typical landmarks include the edge of the tumor, deep grooves of sulci, corners of sulci, convex points of gyri and the horns of the lateral ventricles. After *Rater 1* defined the landmarks in the T2w FLAIR MRIs as the references, *Rater 1* and *Rater 2* then tagged the corresponding landmarks independently within the corresponding US volumes twice. A 1~2-week interval was ensured between the repetitions. The final landmarks in both training and testing database were provided as the averaged results of two trials of landmark marked by both raters (four 3D points for each landmark). The details of the landmarks are listed in Table I for the testing datasets. Similar details for the training dataset can be found in the original publication for the RESECT database [9]. For both sets, a wide range of brain shifts measured as mean initial distances between corresponding landmarks were included to properly examine the performance of registration algorithms.

We employed the mean Euclidean distance between two sets of corresponding landmark points for each patient to assess the intra- and inter-rater variability. For intra-rater variability, we calculated the metric between two trials of landmark picking for each rater; for inter-rater variability, the average of two trials for each rater was first computed and used to obtain the value between two raters. The intra- and inter-rater variability evaluations are presented in Table II for both training and testing data.

TABLE II
INTER- AND INTRA- RATER EVALUATIONS WITH MEAN EUCLIDEAN DISTANCE BETWEEN LANDMARK SETS

| Type | Intra-rater Rater 1 | Intra-rater Rater 2 | Inter-rater R1 vs. R2 |
|---|---|---|---|
| Training data | 0.47±0.10 mm | 0.33±0.06 mm | 0.33±0.08 mm |
| Testing data | 0.21±0.10 mm | 0.48±0.22 mm | 0.42±0.17 mm |

The results are shown as mean±standard deviation.

### III. CHALLENGE SETUP

The CuRIOUS2018 challenge started on April 1$^{st}$, 2018 when the challenge website went live on curious2018.grand-challenge.org. In the next few days, several groups who were active in the field of MRI-iUS registration were identified by literature search and were invited to participate. The challenge was also widely advertised on mailing lists and on bulletin board of medical imaging conferences held in the first half of 2018. Another factor that leads to a good participation was the incentive of generous support of challenge sponsors, which provided a total of 2,100 € for the top three winners. The challenge consisted of two phases.

In phase I, all the teams were required to submit a short paper that elaborated the technique and results on the 22 patients in the RESECT database. These papers were then peer-reviewed and the final camera-ready conference papers were submitted in July 2018.

Phase II started in August 2018, when all the participants who had submitted reports and results on the training data were provided with MRI and iUS data from 10 additional patients (test data). These datasets had identical data curation and annotation protocols as the RESECT database. The location of landmarks in the MRI was provided to the teams, and the teams had to return the locations of those landmarks after MRI-iUS registration within 13 days of the data release. All teams presented their methods and results on the training data at the challenge event, which took place in conjunction to MICCAI 2018 in Granada, Spain.

The RESECT database remains public, and has been downloaded 267 times since its release in April 2017. The test datasets were only released to the participants and the locations of the ground truth landmarks in these datasets remains private. The organizers will continue the challenge in 2019 by adding iUS test data collected during and after tumor resection.

### IV. EVALUATION

The evaluation metric and ranking system are key criteria for the success of a challenge. The metric should reflect the overall quality of the methods and the ranking system should be as fair as possible. It is worth noting that our evaluation method was published on the official website before the challenge took place and was not modified afterwards. Although such transparency in the evaluation process may seem obvious, [18] reported that this transparency was not guaranteed in about 40% of biomedical challenges, which could lead to controversy.

The first component of the evaluation process is a metric to assess the quality of the registration methods. More than 80% of the tasks in biomedical challenges concern segmentation, with the Dice similarity coefficient as the most common evaluation metric [18]. However, challenges with image registration, especially from different modalities, are rarer and we could not find any standard metrics from these competitions. We thus chose to rely solely on the expert-labeled anatomical landmark pairs, by computing the Euclidian distances between the transformed MRI landmarks, after registration, and the ground-truth landmarks defined in the iUS images.

The second component of the evaluation process concerns how the results for each test case are aggregated to rank the teams. The two main options are 1) aggregate the results on all test cases, then rank; or 2) rank by test case, then aggregate the ranks. In the first scenario, we would have ranked the teams based on the mean distance computed from all landmarks of all cases. Instead, we chose the second scenario because it is better fitted to handle missing cases. For each case, we also ranked fully-automatic methods over semi-automatic methods. To aggregate the case-by-case ranks, we simply computed the mean rank of each team.

The evaluation system was as follows:
1. For each test case and for each team, compute the Euclidian distances between landmark pairs after registration, i.e. between the transformed MRI landmarks and the ground truth iUS landmarks.
2. For each test case, rank teams according to their mean distance between landmark pairs. Exceptions include:
    a. If one team could not provide results for a test case, or if these results could not be processed for any reason, then that team is ranked last for the test case.
    b. If two mean distances differ by less than 0.5 mm, a team with a fully-automatic method is ranked higher than a team with a semi-automatic method.
3. Compute the mean rank of every team, which gives the final ranks of the Challenge.

### V. CHALLENGE ENTRIES

#### A. Team cDRAMMS

Machado et al. [13] extended the Deformable Registration via Attribute Matching and Mutual-Saliency Weighting (DRAMMS) algorithm [19], a general-purpose algorithm [20], specifically for the US-MRI registration problem, which they termed as correlation-similarity DRAMMS or cDRAMMS. They released it at https://www.nitrc.org/projects/dramms/ (version 1.5.1). The original DRAMMS has two good properties for US-MRI registration. First, representing each voxel with multi-scale and multi-orientation Gabor attributes in DRAMMS offers a richer information than purely image intensities. This helps to establish more reliable voxel correspondences despite the different image protocols and different intensity profiles between US and MRI images. Second, the mutual-saliency module in DRAMMS automatically assigns low confidence or weights to regions that cannot establish reliable or cannot find counterparts across images. This potentially reduces the negative effects of the missing correspondences between US and MRI images. Different from the original DRAMMS, which uses the sum of square differences (SSD) between attributes for matching, the modified cDRAMMS uses correlation coefficient [21] and correlation ratio [22] on attributes for voxel matching. CC and CR on voxel attributes in cDRAMMS establish voxel correspondences at a higher accuracy and higher reliability than SSD in DRAMMS.

### B. Team DeedsSSC

Heinrich et al. [14] used DeedsSSC, which comprises a linear and a non-rigid registration that are both based on discrete optimization and modality-invariant image features. Specifically, self-similarity context features (SSC) are extracted for both MRI and ultrasound scans that are matched based on a dense displacement sampling. First, the similarity maps for each considered control point are used to extract correspondences for fitting a linear transform using least trimmed squares, similar as done in block-matching approaches. Second, new similarity maps are calculated for linearly aligned images and an efficient graphical model based discrete optimization (deeds) is used to estimate a nonlinear displacement field that avoids implausible warps and further improves the registration quality. All computations are performed for scans resampled to isotropic 0.5 mm resolution and using the default parameters (see https://github.com/mattiaspaul/deedsBCV) with an optimization over multiple grid-scales. Finally, the nonlinearly warped landmarks are again constrained to follow a rigid 6-parameter transform for improved robustness. The algorithm is executed within less than 10 seconds per scan pair on a multi-core CPU and ongoing work considers the huge potential for further speed-ups through parallelized GPU computations.

### C. Team FAX

Zhong et al. [17] proposed a learning-based approach to resolve intraoperative brain as an imitation game. This point-based approach predicts the deformation vectors of key points to compensate the non-rigid brain-shift. For each key point, they extract a local 3D patch in iUS and model the key point distribution as the encoding of the current observation. A demonstrator is constructed providing the optimal deformation vector based on the current key point location and the ground truth. An artificial neural network is trained to imitate the behavior of the demonstrator and to predict the optimal deformation vector given current observation. To increase robustness, the proposed technique uses a multi-tasking network with a rigid transformation as auxiliary output. In addition, we use a non-rigid deformation to augment the 3D volume and 3D key points to facilitate the training.

### D. Team ImFusion

The method [16] is based on the multi-modal similarity metric $LC^2$ [7] and has recently been used in a first live evaluation during surgery [23] (data NOT overlapping with challenge data). A non-linear optimization algorithm changes the values of a parametric transformation model to maximize it. In a pre-processing step specific to the challenge data set (cartesian 3D ultrasound volumes compounded by the SonoWand system), the volume sides facing the ultrasound probe is estimated and the outermost 4mm of content are cropped accordingly. The registration algorithm is implemented in the proprietary ImFusion SDK with full OpenGL-based GPU acceleration. The ultrasound volume is assigned as fixed volume, resampled to 0.5mm (half the MRI voxel size), and properly zero-masked. The chosen similarity metric patch-size is $7 \times 7 \times 7$ voxels, as optimized in prior work. Two non-linear optimizers successively operate on the parameters of a rigid pose from the initialization as provided by the navigation system. The first is a global DIRECT (DIviding RECTangles) sub-division method [24] searching on translation only, followed by a local BOBYQA (Bound Optimization BY Quadratic Approximation) algorithm [25] on all six parameters. The local optimizer then executes another search on full affine parameters in order to accommodate non-uniform scaling and shearing of the data.

### E. Team MediCAL

Multimodal deformable registration between the MRI and intra-operative 3DUS was achieved with a weighted version of the locally linear correlation metric ($LC^2$), correlating MRI intensities and gradients with ultrasound, while adapting both hyper-echoic and hypo-echoic regions within the cortex. The method [15] was initialized with a global rotation of the US volume to match the orientation observed on the MRI. This was achieved using a PCA of the extracted inferior skull region, identifying the principal orientation vectors of the head, followed by a scaling and translation correction. This fusion step uses a patch-based approach of the US voxels, comparing intensity and gradient magnitudes extracted from the MRI with a linear relationship. The registration applies sequentially a rigid and non-rigid step, with the later integrating a weighting term and controlled by a cubic $5 \times 5 \times 5$ B-Spline interpolation grid, distributed uniformly in the fan-shaped US volume. The weighting term uses pre-annotated labels on the MRI, representing both the hypoechoic (fluid cavities) and the hyperechoic (ex. choroid plexus) areas observed on ultrasound. This term is added only at the non-rigid step as it is highly specific to the internal areas in US such as the lateral ventricles, requiring a rigid pre-alignment. Registration optimization was performed using BOBYQA, which avoids computing the metric's derivatives.

## F. Team NiftyReg

Drobny et al. [12] suggest a method which uses a block-matching approach to automatically align the pre-operative MRI with the iUS image. The registration algorithm used is part of the NiftyReg open-source software package [26]. The block-matching method of the registration stage iteratively establishes point correspondences between the reference image and the warped floating image and then determines the transformation parameters using least trimmed squares (LTS) regression. A two-level pyramidal approach for coarse to fine registration is used. For the block-matching, both images are divided into uniform blocks of 4 voxel edge length. The 25% of blocks with the highest intensity variance in the reference image are used and the rest are discarded. Each of these image blocks is compared to all floating image blocks that overlap with at least one voxel. The floating image matching block for each reference block is determined as the one with maximum absolute normalized cross-correlation (NCC). After establishing the point-wise correspondences the second step is the update of transformation parameters via LTS regression. At every iteration, the composition of the block-matching correspondence and the transformation of the previous step determines the new transformation by LTS regression.

## VI. RESULTS

### A. Phase I: distances on the training data

The results obtained on the training dataset were reported by each team in their respective contribution to the challenge proceedings [27]. Most authors reported their distances obtained after registration for each case, although some reported only averaged values. Table III summarizes the mean distance between landmark pairs after registration, over all landmark of all cases, computed by each team. All teams but one improved from the initial distances, with three teams achieving a mean distance under 1.75 mm and two more under 3.35 mm. Team cDRAMMS initial reported a mean distance between landmarks of $3.35 \pm 1.39$ mm. With an updated version of their method, this error was later reduced to $2.28 \pm 0.71$ mm. Sun et al. [11] provided partial results on 4 cases only, since the other 18 cases were used to train their neural network. This team eventually did not participate to the second phase.

TABLE III
SUMMARY OF THE CHALLENGE RESULTS.

| Team | Distances between landmark pairs after registration Mean ± std, in mm Training set / Test set | | Mean case-by-case rank | Final challenge rank |
|---|---|---|---|---|
| cDRAMMS | 3.35 ± 1.39 | 2.18 ± 1.23 | 3.4 | 3 = |
| DeedsSSC | 1.67 ± 0.54 | 1.87 ± 0.93 | 2.4 | 2 |
| FAX | 1.21 ± 0.55 | 5.70 ± 2.93 | 5.3 | 5 = |
| ImFusion | 1.75 ± 0.62 | 1.57 ± 0.96 | 1.5 | 1 |
| MedICAL | 4.60 ± 3.40 | 6.59 ± 2.89 | 5.3 | 5 = |
| NiftyReg | 2.90 ± 3.59 | 3.21 ± 3.57 | 3.1 | 3 = |
| Hong et al. | 5.60 ± 3.94 | 6.65 ± 4.55 | - | - |
| *Sun et al. | 3.91 ± 0.53 | - | - | - |
| Initial distances | 5.37 ± 4.27 | 6.38 ± 4.36 | - | - |

For each team, the first columns give the mean distances between landmark pairs after registration, computed over all landmark of all cases, for the training and test sets. The mean case-by-case rank, computed on the test set only, and the final challenge rank are then given. For comparison, the last line contains the mean initial distances, before registration. Teams cDRAMMS and NiftyReg were eventually ranked tied at third (=). Hong et al. sent results on the test data but did not attend the challenge event. Sun et al. sent only partial results (*) on the training set, but did not participate to the second phase of the challenge. These two teams were thus not ranked.

### B. Phase II: distances on the test data

This section presents the results of the 6 teams that completed phase II of the challenge, on the test dataset. Figure I first shows the results per test case, aggregated across all teams. Test cases with the largest initial error (cases 1, 5, and 3) were the most difficult to treat. Results for test cases with the smallest initial error (4 and 9) were in average improved, also several teams obtained larger distances after registration. Finally, results were consistently improved for all other cases with an initial error in the 4-6 mm range.

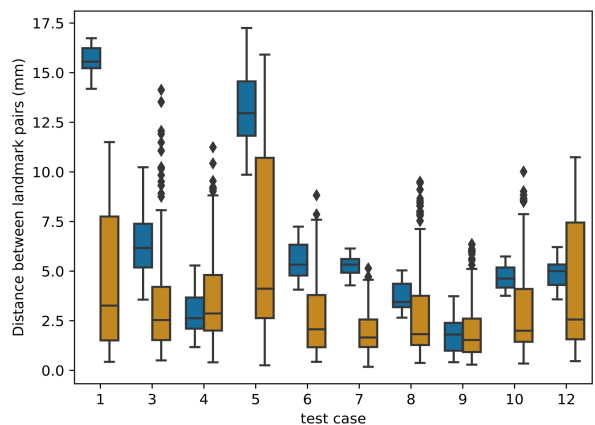

Figure I. Results per test case: box plot distribution of the distances between landmark pairs. For each test case, the left box plot (blue) shows the initial distances before registration. while the right box plot (orange) shows the distribution after registration, aggregated over all teams.

Regarding team-by-team results, mean distances between landmark pairs after registration are summarized in Table III while the distribution of these distances is detailed in Figure II. Teams ImFusion and DeedsSSC obtained a mean distance between landmark pairs well below 2 mm, respectively of 1.57 and 1.87 mm. These excellent results are consistent across all test cases, with a standard deviation around 1 mm for both teams, which confirmed the results reported on the training data set. Team cDRAMMS also consistently obtained very good results, with a mean error of 2.18 mm and a single large residual error of 4.3 mm for case 5. Results of team NiftyReg are more contrasted. As can be seen on the lower panel of Figure II, they obtained excellent results for all cases but two, cases 1 and 5, where the distance was only reduced from 15.7 to 5.9 mm and 13.2 to 12.8 mm, respectively. Without these two outliers, the mean distance over all cases would be reduced from $3.21 \pm 3.57$ mm to $1.70 \pm 0.91$ mm. Team FAX reported the best results on the training set, with a mean distance between landmark pairs of $1.21 \pm 0.55$ mm. However, this distance leaped to $5.70 \pm 0.55$ mm on the test data, which potentially shows their deep learning method overfitted the data during the training phase. Finally, team MedICAL obtained few or no improvements from the initial distances between landmark pairs.

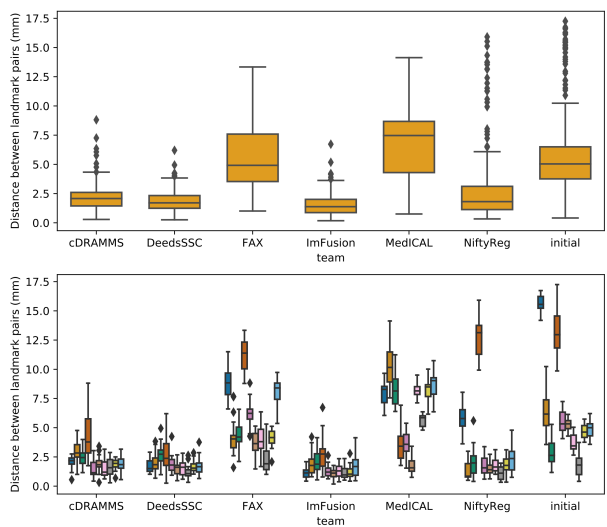

Figure II. Distribution of the distances between landmark pairs obtained by each team after registration, on the test set. For comparison, the last column contains the initial distances before registration. The upper panel shows the global results computed over all landmarks of all test cases. In the lower panel, these results are split by test case.

*C. Phase II: complementary criteria*

All methods were fully automatic. Although it was not a factor in the evaluation, several teams reported their computation time. These values range from 1.8 sec for team FAX (architecture not specified), to approximately 20 sec for both DeedsSSC and ImFusion on laptops, and to 103 sec for team MedICAL. Computation time of teams cDRAMMS and NiftyReg were not provided.

*D. Qualitative results*

To demonstrate the data and the registration task, MR and iUS volumes of one patient was chosen from each of the training and test datasets, and are shown in Figures III and IV, respectively. The selected cases have a relatively large initial mTRE and a substantial variability between the teams. Also, note that these two cases do not necessarily directly reflect the overall ranking of the challenge, which was based on averaged rankings of all cases. As no quantitative measures of registration quality are available in a clinical setting, visual inspection of the images is important to obtain an impression of the registration quality. As shown in Figures III and IV, the registration accuracy can be evaluated by adapted visualization and identification of homologous features such as sulci, gyri and ventricles in the images.

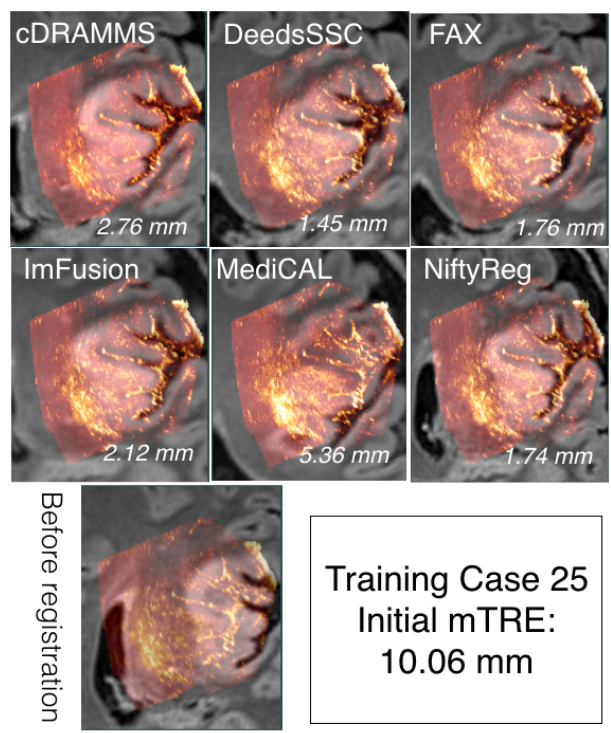

Figure III. Qualitative comparison of registration results for Training Case 25 across different teams. For each team, the ultrasound and deformed FLAIR MRI scan is overlaid together. The mTRE values for each team is listed at the right bottom corner of each image overlay.

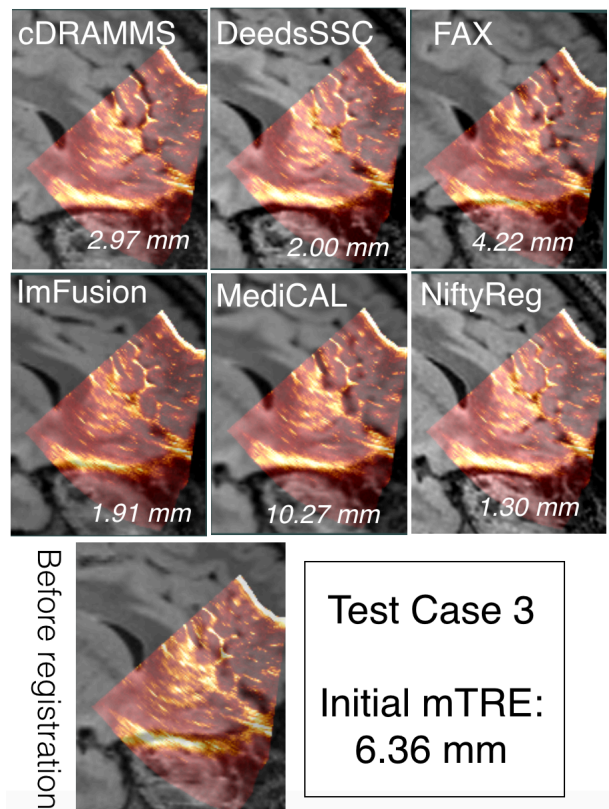

Figure IV. Qualitative comparison of registration results for Test Case 3 across different teams. For each team, the ultrasound and deformed FLAIR MRI scan is overlaid together. The mTRE values for each team is listed at the right bottom corner of each image overlay.

*E. CuRIOUS2018 Challenge ranks*

Following the description in Section IV, all teams were ranked independently for each test case, based on the mean distances between landmark pairs. These case-by-case ranks are summarized in Figure V, with the number of times each team was ranked at the $i^{th}$ place, with $i$ from 1 to 6.

The winner and runner-up are teams ImFusion and DeedsSSC, which are perfectly consistent with their respective results reported in Figure II. Note that ImFusion obtained the best registration for 6 of the 10 tests cases. Despite a larger mean registration error, team NiftyReg was ranked third before team cDRAMMS as it obtained a better case-by-case rank (3.1 vs 3.4). However, team cDRAMMS also had very good results but consistently handled all cases, including the extreme ones. This specific situation pointed out the fact that the challenge metric favors accuracy over precision, with a limited penalty when low quality results are obtained on a single case or two. To overcome this limit, as we consider precision is a crucial factor for the surgeons' acceptance of a method, the challenge's organizers decided to declare a tie for third place. Both NiftyReg and cDRAMMS thus received the same third place prize. Finally, both teams FAX and MedICAL obtained a mean case-by-case rank of 5.3, and were ranked tied at the $5^{th}$ place in the challenge.

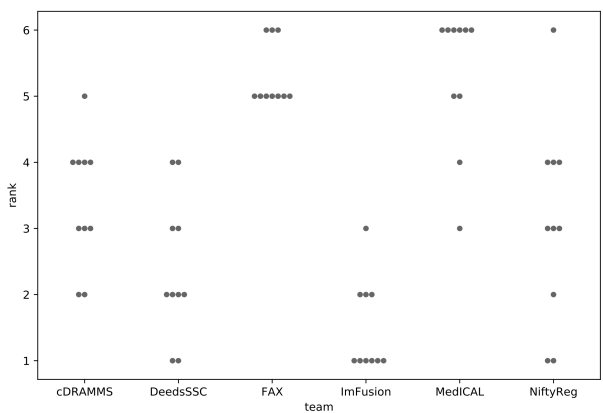

Figure V. Case-by-case ranks for each team. For example: over the ten test cases, team ImFusion was ranked first six times, second three times, and third one time.

VII. DISCUSSION

In this challenge, the focus has been on MR-iUS registration in the context of brain tumor surgery. As both the training and test datasets exclusively contain data from LGG surgeries, there has been a special focus on this tumor type. The resection of LGGs is particularly challenging as the tumor tissue can be very similar to normal brain tissue. In LGG surgery, there are also fewer options for additional guidance as tools like 5-ALA fluorescence are not available. Intraoperative ultrasound is therefore an attractive solution in these cases. The optimization and benchmarking of available registration algorithms on data from these tumors is therefore particularly important for successful future clinical translation. Even though the emphasis has been on LGG, the results from the challenge will generalize well to other tumor types such as HGGs and metastasis as these tumors are more distinct from normal brain tissue and depict clearer boundaries than LGGs in ultrasound images.

An important obstacle for the widespread use of iUS is the challenging and unfamiliar image interpretation. The integration of iUS into the navigation system and the visualization of corresponding slices in pre-operative MR and iUS makes this interpretation considerably more intuitive. With accurate MR-iUS registration, the surgeon can perform the resection based on the MR images even after brain shift which makes the neuronavigation accurate and easy to interpret. MR-iUS registration also enables correction of other types of pre-operative MR data such as fMRI and DTI [28].

Image registration techniques tailored for MRI-iUS registration in this challenge were landmark-, intensity- or learning-based. The performance of landmark-based methods in non-linear image registration depends on both finding enough landmarks that cover the entire volume, and correctly finding their corresponding landmarks in the second volume. The voxel-wise attribute-based method of Machado et al. [13] (team cDRAMMS) did relatively well despite the fact that iUS and MRI have drastically different salient features, and ranked third in a tie with Drobny et al. [12] (team NiftyReg). The top three algorithms in this challenge [12-14] were all intensity-based techniques, which calculated a dense transformation map by utilizing intensity values at all locations.

Deep learning has had a large success in segmentation and classification problems in medical image analysis, but its success in image registration has been much less impressive [29]. The two submissions that used DL in this challenge were from Sun et al. [11], who did not participate to the second phase, and Zhong et al. [17] (team FAX), who ranked first in the results reported on the training database. However, their method did not work well on the test database. A common culprit for such behavior is overfitting, where the model overfits the training data and therefore performs poorly on the unseen test data. As more training data becomes available, this method is expected to perform better in the future.

Symmetric image registration techniques provide unbiased estimates of the transformation field and are known to generally outperform their asymmetric counterparts [30]. Two of the top three methods in this challenge [12, 14] compared the performance of their techniques in symmetric and asymmetric settings. They both concluded that asymmetric transformations lead to a superior performance in this challenge. This is an interesting finding and is likely due to the vast differences in physics of US and MR imaging modalities.

It was noticed by some of the challenge participants and discussed during the event that in some cases affine transformations outperformed non-linear elastic transformations. This might seem surprising as brain shift is often described as a non-uniform deformation. However, before resection a large component of the experienced mismatch between MRI and iUS is often due to inaccurate patient-MRI registration. This is a rigid registration most often based on anatomical landmarks, fiducials, surfaces or a combination of these. Consequently, an affine transformation might be sufficient to correct for most of the misalignment. After resection, the situation will be different with larger and highly non-linear deformations and affine transformations will likely not be sufficient to register the images.

In both the training and test databases, we selected landmarks that cover a large part of the iUS volume with maximal distance between neighboring landmarks. This strategy provides a good benchmark for comparison of image registration techniques. However, the quality of the alignment closer to the tumor is more clinically important as it better helps the neurosurgeon to optimize the resection size and location.

The distance between corresponding landmarks in the two images before and after registration is a well-established metric for evaluation of registration results in the absence of a ground truth. Despite being widely used, this metric has some limitations. As there is only a limited number of landmarks associated with each image, the registration error is only evaluated at a limited number of locations and will therefore not capture local displacements and deformations in other locations. The number of landmarks and their distribution in the image volume are therefore important. The landmarks in both the training and test sets have therefore been carefully placed in order to capture the displacements and deformations as good as possible. However, we noticed that in test case #5 the registration results were not accurate by visual inspection even though the mTREs indicated successful alignment. This emphasizes the need for both quantitative and qualitative assessment of registration results. Another limitation of this metric is the localization error associated with manual placement of points. Also, for the landmarks to be valid for evaluation of registration results, this localization error has to be significantly lower than the expected registration errors. We have measured the inter- and intra-rater variability in both the training and test data and shown that these are significantly lower than the registration errors. Even though the landmarks do not represent the absolute ground truth, they are valid for the evaluation of the registration results. The use of landmarks as the only metric for the challenge also represent a limitation as other important characteristics such as computation time are not measured. For implementation in a clinical setting, for example, other characteristics would also be of critical importance. However, in the challenge setting, the use of a single well-defined metric is advantageous. A single metric enables a straightforward, comprehensible ranking scheme and an open, fair competition. With the use of multiple metrics, there will always be a discussion of the weighting of the different characteristics and how to aggregate the results. The rules for aggregation of the ranks in this challenge were outlined before the challenge and were not changed at any point. Still, the system used favors accuracy over precision. As discussed during the challenge event, for the clinical users, high precision and high accuracy are equally important and precision can even be more important than accuracy. This point should be re-designed and improved in future editions of the challenge.

## VIII. FUTURE WORK

In the first edition, the registration task solely focused on MRI-iUS registration before dura-opening and after craniotomy. However, with the progress of tumor resection, tissue deformation is an on-going process, and accurate tracking can ensure the complete removal of cancerous tissues, preventing any additional surgeries. Intended as a recurrent open challenge to further improve the registration algorithms, we expect to introduce multiple sub-challenges in future CuRIOUS challenges to target brain shift correction at different stages of the surgery, especially during and after resection.

For clinical practices, besides accuracy and robustness, processing speed is an imperative factor. In the inaugural edition of the challenge, performance speed was not emphasized in scoring the teams because it can be affected by multiple factors, including implementation platforms, for prototype algorithms. In future challenges, we aim to place discussions and emphasis on this topic, as well as optimization algorithms to direct the results of the challenge towards more realistic clinical implementations.

## IX. CONCLUSION

Holding great clinical values, MRI-iUS registration for correcting tissue shift in brain tumor resection is still a difficult task. As the first public image processing challenge to tackle this clinical problem, the CuRIOUS2018 Challenge provided a common platform to evaluate and discuss existing and emerging registration algorithms on this topic. The results of CuRIOUS2018 provided valuable insights for the current developments and challenges from both the technical and clinical perspectives. This is an important step forward to help translate research-grade automatic image processing into clinical practice to benefit the patients and clinicians.

ACKNOWLEDGMENT

The work is supported by Norwegian National Advisory Unit for Ultrasound and Image guided Therapy, NSERC Discovery RGPIN-2015-04136. The organizers would like to thank Medtronic and the French ANR within the *Investissements d'Avenir* program under references ANR-11-LABX-0004 (Labex CAMI) for their generous sponsorship. The work for the DeedsSSC algorithm was funded in part by the German Research Foundation (DFG) under grant number 320997906 HE 7364/2-1. Drobny et al. were supported by the UCL EPSRC Centre for Doctoral Training in Medical Imaging and Wellcome/EPSRC Centre for Interventional and Surgical Sciences [NS/A000050/1], and the Wellcome/EPSRC Centre for Medical Engineering [WT 203148/Z/16/Z] and EPSRC [NS/A000027/1].